\documentclass[11pt,twoside]{article}
\usepackage{asp2004}
\usepackage{epsf}
\usepackage{psfig}
\usepackage{lscape}
\usepackage{graphicx}

\begin{document}
\title{A Grid of FASTWIND NLTE Model Atmospheres of Massive Stars} 

\author{K.~Lefever$^1$, J.~Puls$^2$, C.~Aerts$^{1,3}$}
\affil{$^1$ Instituut voor Sterrenkunde, Katholieke Universiteit Leuven,
Celestijnenlaan 200D, B-3001 Leuven, Belgium} 
\affil{$^2$ Universit\"atssternwarte M\"unchen, Scheinerstrasse 1, D-81679 
M\"unchen, Germany}
\affil{$^3$ Departement Astrofysica, Radboud Universiteit Nijmegen, PO
  Box 9010, 6500 GL Nijmegen, the Netherlands}

\begin{abstract}
In the last few years our knowledge of the physics of massive stars has
improved tremendously. However, further investigations are still needed,
especially regarding accurate calibrations of their fundamental parameters.
To this end, we have constructed a comprehensive grid of NLTE model
atmospheres and corresponding synthetic spectra in the massive star domain.
The grid covers the complete B type spectral range, extended to late O on
the hot side and early A on the cool side, from supergiants to dwarfs and
from weak stellar winds to very strong ones. It has been calculated with the
latest version of the FASTWIND code. The analysis of an extensive sample of OB
stars in the framework of the COROT space mission will lead to accurate
calibrations of effective temperatures, gravities, mass loss rates etc. This
paper contains a detailed description of the grid, which has been baptised as
BSTAR06 and which will be available for further research in the near future.
\end{abstract}

\section{Introduction}

In the last decade, with the advent of high-resolution, high signal-to-noise
spectroscopy, massive stars, and the many astrophysical processes influenced
by them, have regained considerable interest. Significant improvements have
been achieved in the development and accuracy of models predicting the
stellar atmospheres and their winds. Still, it is surprising to note how many
things we do not understand, just to name, e.g., the formation of structure
(loosely called ``clumping'') in the stellar wind and its consequences. Though
a lot of progress has been made in the O and A type regime, the
B type regime in between still suffers from many shortcomings.  

Since the detailed spectroscopic analysis of individual objects is a rather
time-consuming (and boring!) job, scientists refrained from analysing large
samples and never tackled more than a few tens of stars at once. Therefore
and unfortunately, knowledge in the B type regime still relies on small
number statistics and could thus only gain from large sample studies. As an
alternative to (automatic) high precision analyses (e.g., by means of
genetic algorithms, \citealt{Mokiem2005}, or neuronal networks) one could
profit from a grid-method, which offers a good compromise between
effort, time and precision when an {\it appropriate grid} has been set up.

Since one of our goals is to derive effective temperature calibrations for
the complete B type spectral range, for stars in different evolutionary
phases, in a way as homogeneous as possible, we need both a comprehensive 
sample of stars {\it and} a huge grid of reliable model atmospheres. The
first has been supplied to us by the COROT team, and consists of a database
of ground-based observations, obtained in preparation of the space mission 
and for which they need accurate fundamental parameters. The database
comprises FEROS, ELODIE and SARG spectra of some 350 massive OBA stars
brighter than 9.5 mag. To satisfy our second need, we developed an
appropriate and extensive grid of model atmospheres in the way discussed in
Section 3.

\section{Input Physics and Line Profiles}

BSTAR06 is a grid of NLTE, line-blanketed, unclumped model atmospheres,
calculated by means of the recent version of the atmosphere code FASTWIND.
From this grid we want to retrieve, within a minimum of time, the
fundamental parameters of huge samples of hot massive stars with winds. This
will be done by comparing synthetic and observed spectra in a way
which is as automated as possible. One can follow the evolution of the code
through the years in the papers by \citet{SantolayaRey1997},
\citet{Herrero2002} and \citet{Puls2005}. We refer to the latter for a
detailed description of the code and the involved input physics. 

So far, we have calculated selected optical and IR line profiles for the
elements H, He and Si, since these provide appropriate and sufficient
diagnostics to derive the stellar and wind parameters of B type stars.  In
particular, the different ionisation stages of silicon (i.e., Si~II, Si~III and
Si~IV) will be used to fix the effective temperature T$_{\rm eff}$. Once we know
the latter, we can easily derive the surface gravity, log $g$, from the wings of
the Balmer lines. Since H$\alpha$ forms further out in the atmosphere than
the other Balmer lines, it is affected by the wind, and enables good
diagnostic for the wind parameters $\dot{M}$ (the mass loss rate),
v$_{\infty}$ (the terminal wind velocity) and $\beta$ (the velocity field
exponent), at least if H$\alpha$ is not too much in absorption (i.e., the
wind is not too weak). The method used to determine
the physical parameters is described in full detail in \citet{Lefever2006}.

\section{Description of the BSTAR06 Grid}

The BSTAR06 grid has been set up to cover the complete parameter space of
B type stars.  As such a grid shall also be a good starting point for a
(follow-up) detailed spectroscopic analysis of massive stars, it has been
constructed as representative and dense as possible within a reasonable
computation time.  

In total, we have calculated 264\,915 models. We consider 33 effective
temperature gridpoints, ranging from 10\,000~K to 32\,000~K, in steps of 500~K
below 20\,000~K and in steps of 1\,000~K above.  In this way we will be able
to deal with all stars with spectral types within early A until late O.

As we will analyse massive stars in different evolutionary stages, from main
sequence up to supergiants, the gravities comprise the range of $\log g$ =
4.5 down to 80\% of the Eddington limit, in steps of 0.1, thus resulting in
a mean number of 28 values at each effective temperature point.

For each ($T_{\rm eff}$, log $g$)-gridpoint, we have adopted one 'typical'
value for the radius, $R_*$, keeping in mind that a rescaling to the `real'
values is required once concrete objects are analyzed. In most cases, the
actual radius can then be determined from the visual magnitude, the 
distance of the star and its reddening. As a first approximation for the
grid, the radius $R_*$ and the mass $M_*$ are determined from interpolation
between evolutionary tracks, so that the grid is fully consistent with
stellar evolution.

The chemical composition has been chosen to be representative for the
typical environment of massive stars. As we consider only H, He and Si
explicitely, we have varied only the helium and silicon abundance, whereas
for the remaining background elements (responsible, e.g., for radiation
pressure and line-blanketing), we have adopted a solar composition,
following \citet{Asplund2005}. For helium, three different abundances have
been incorporated: He/H = 0.10, 0.15 and 0.20 by number. As discussed, e.g.,
in \citet*{Lefever2006}, the silicon abundance in B stars is still subject
to discussion. Depending on sample and method, values range from solar to a
depletion by typically 0.3 dex, both with variations by $\pm$ 0.2 dex.
Therefore, also for silicon three abundance values have been adopted, i.e.,
the solar value ($\log$~(Si/H) = -4.49 by number, \citealt{Asplund2005}) and
an enhancement and depletion by a factor of two, i.e., $\log$~(Si/H) = -4.19
and -4.79.

Since our grid should enable the analysis of stars of different luminosity
class and thus wind-strength, we incorporated seven different values for the
wind-strength parameter, $\log Q$ (cf. \citealt{Puls1996}), with $Q$ =
$\dot{M}/ (v_{\infty}\, R_{\ast})^{1.5}$. As for the radius, we were forced
to assume a 'typical' value for the terminal wind velocity, $v_{\infty}$, in
order to reduce extent of the grid. To this end, terminal wind velocities for
supergiants have been either interpolated from an existing, but rather
crude grid of late O/early B type stars, either estimated from observed values
\citep[amongst others]{Kudritzki2000}. 
For non-supergiants, we used a similar scaling
relation as \citet*{Kudritzki2000}, i.e., $v_{\infty}$ = C $\cdot v_{\rm esc}$
(see their equation 9), but with $C = 2.5$ for $T_{\rm eff} \geq
24\,000K$, an interpolation between 1.4 and 2.5 for $20\,000K < T_{\rm eff}
< 24\,000K$ and an interpolation between 1.0 and 1.4 for lower temperatures.
By fixing $R_{\ast}$ and $v_{\infty}$ in this way, we end up with a wide
spread in mass-loss rates for each predescribed $Q$-value. The wind velocity
law is determined by the $\beta$-exponent, for which we considered 5 values
in the grid: 0.9, 1.2, 1.5, 2.0 and 3.0 for the most extreme cases.
 
Finally, for calculating the NLTE model atmospheres we used a microturbulent
velocity, $v_{\rm micro}$, of 8, 10 and 15 km/s for the temperature regimes
$T_{\rm eff} < 15\,000K$, $15\,000K \leq T_{\rm eff} < 20\,000K$ and $T_{\rm
eff} \geq 20\,000K$, respectively, whereas for {\it all} synthetic line profiles
(from all models) microturbulent velocities of 6, 10, 12, 15 km/s have been
used, with an additional value of 3 km/s for $T_{\rm eff} \leq 20\,000K$ and
20 km/s for $T_{\rm eff} > 20\,000K$.

\section{Grid Analysis: Some Representative Figures}

\subsection{Models Located in the HRD}

\begin{figure}[!ht]
\centering \vspace{-0.2cm}
\includegraphics[width=12cm, height=8cm]{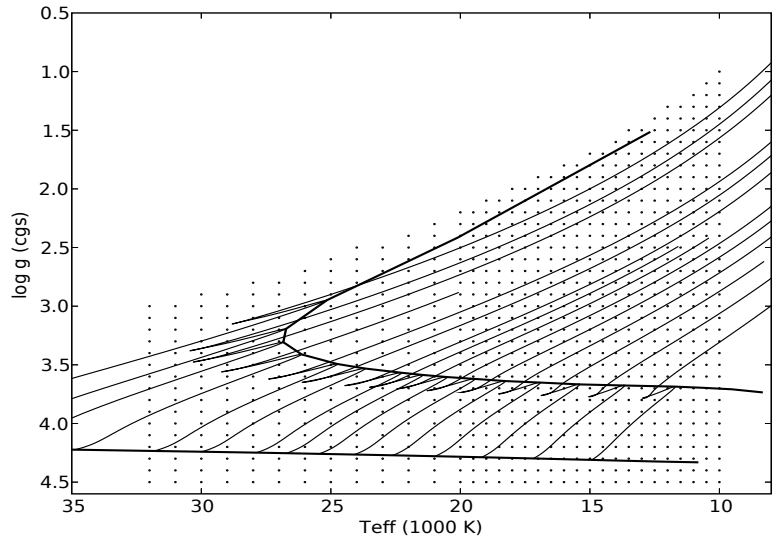}\vspace{-0.5cm}
\caption{Location of the BSTAR06 grid models (dots) in the
($T_{\rm eff}$, log $g$)-plane, compared to standard evolutionary models
without taking into account rotation and a stellar wind  
\citep{Pamyatnykh1999}, starting at the right at an intial stellar mass of 4
M$_{\odot}$, until 40 M$_{\odot}$ at the left. The ZAMS and TAMS are
indicated by thick solid lines.}
\label{HRD}
\end{figure}

In Fig.\ref{HRD} we display the position of our models in the
Hertzsprung-Russell Diagram (HRD), in comparison with evolutionary tracks of
stars with initial stellar masses between 0.4 and 40 M$_{\odot}$. Obviously, 
we completely cover the evolutionary sequences of these stars, from MS to SG
phase, within the temperature range of the B type stars.

\subsection{Diagnostic Lines and their Isocontours of Equivalenth Width}

\begin{figure}[!ht]
\includegraphics[scale=0.35]{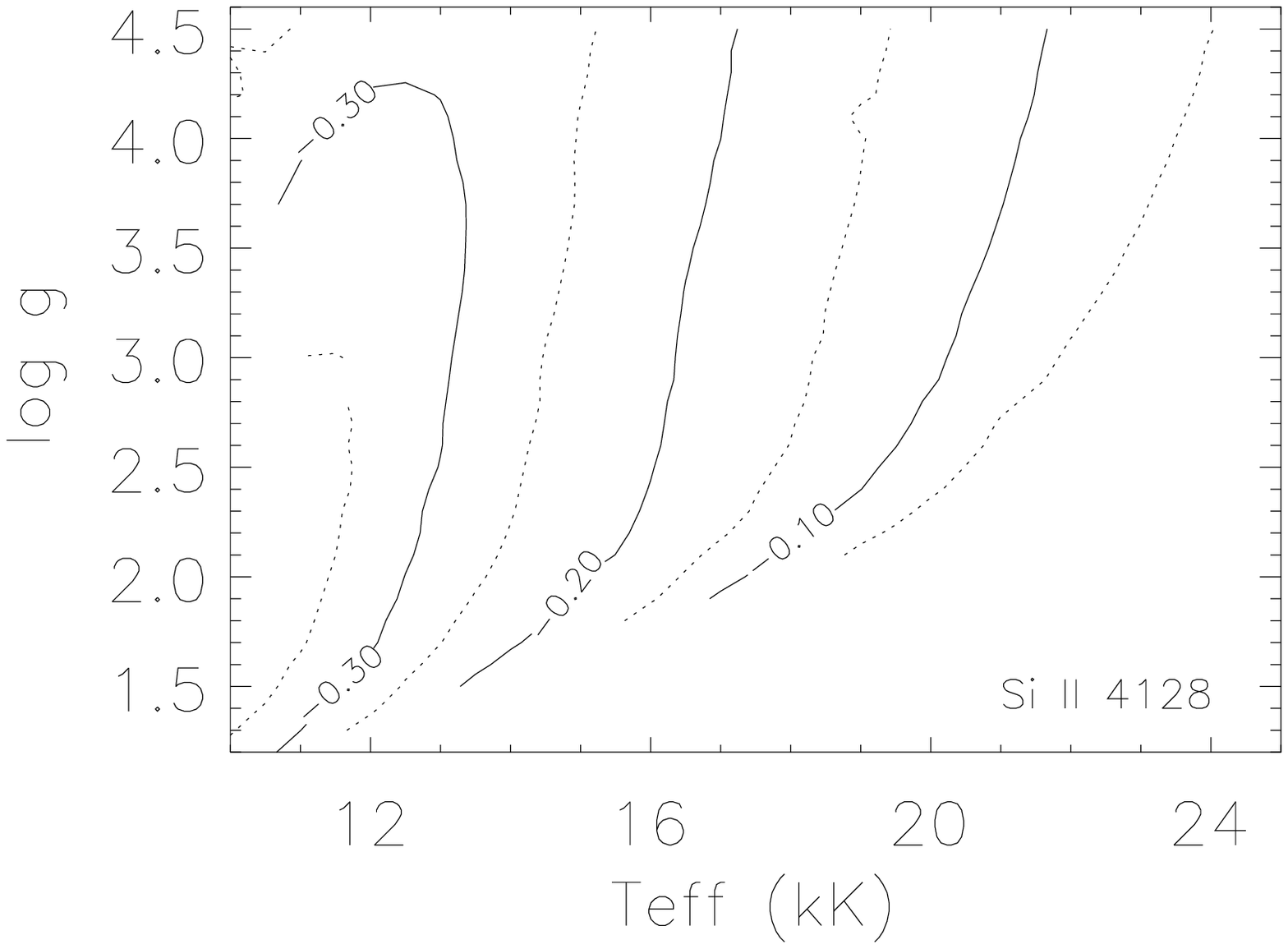}\hspace{-0.5cm}
\includegraphics[scale=0.35]{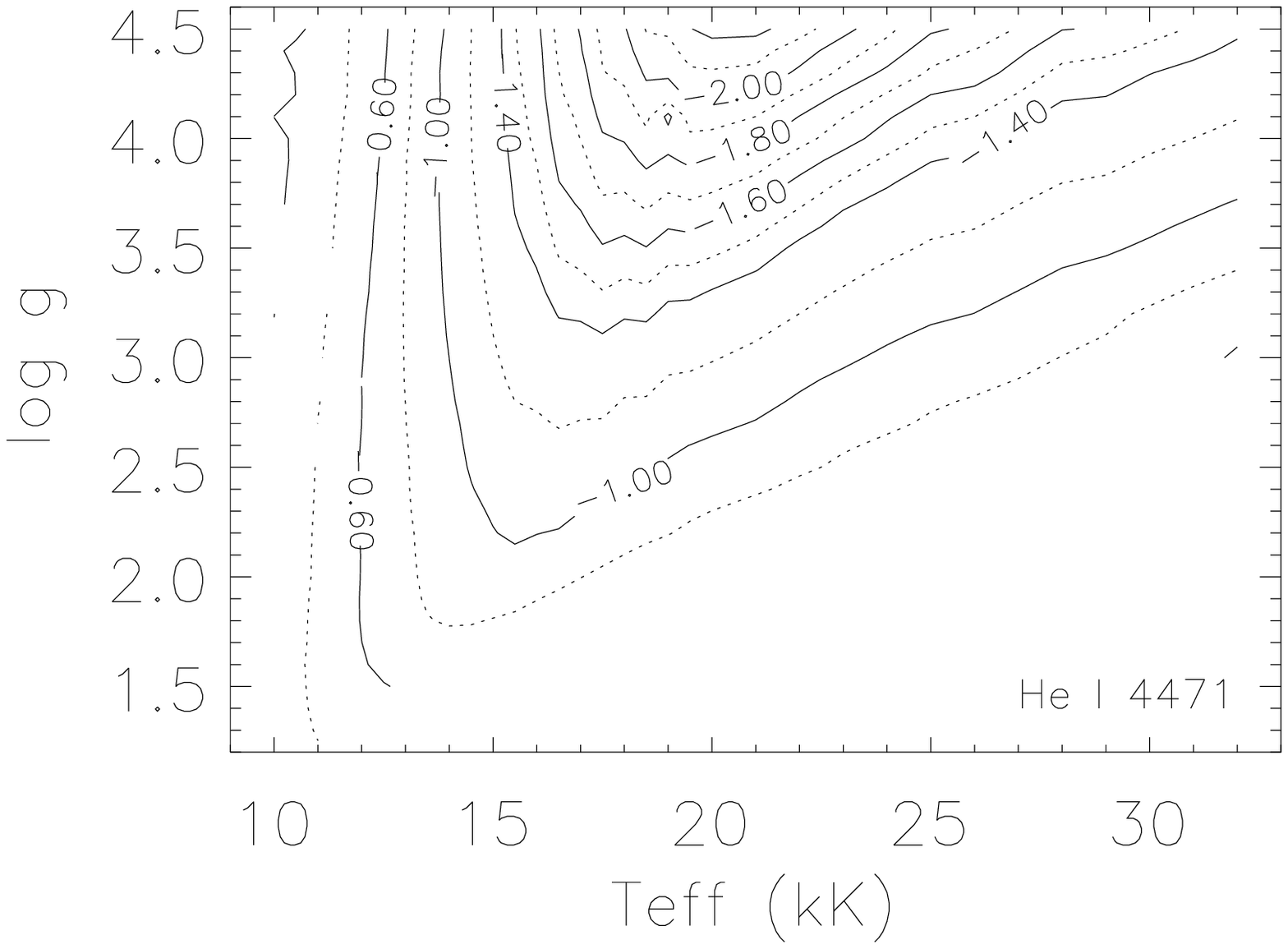}\\
\includegraphics[scale=0.35]{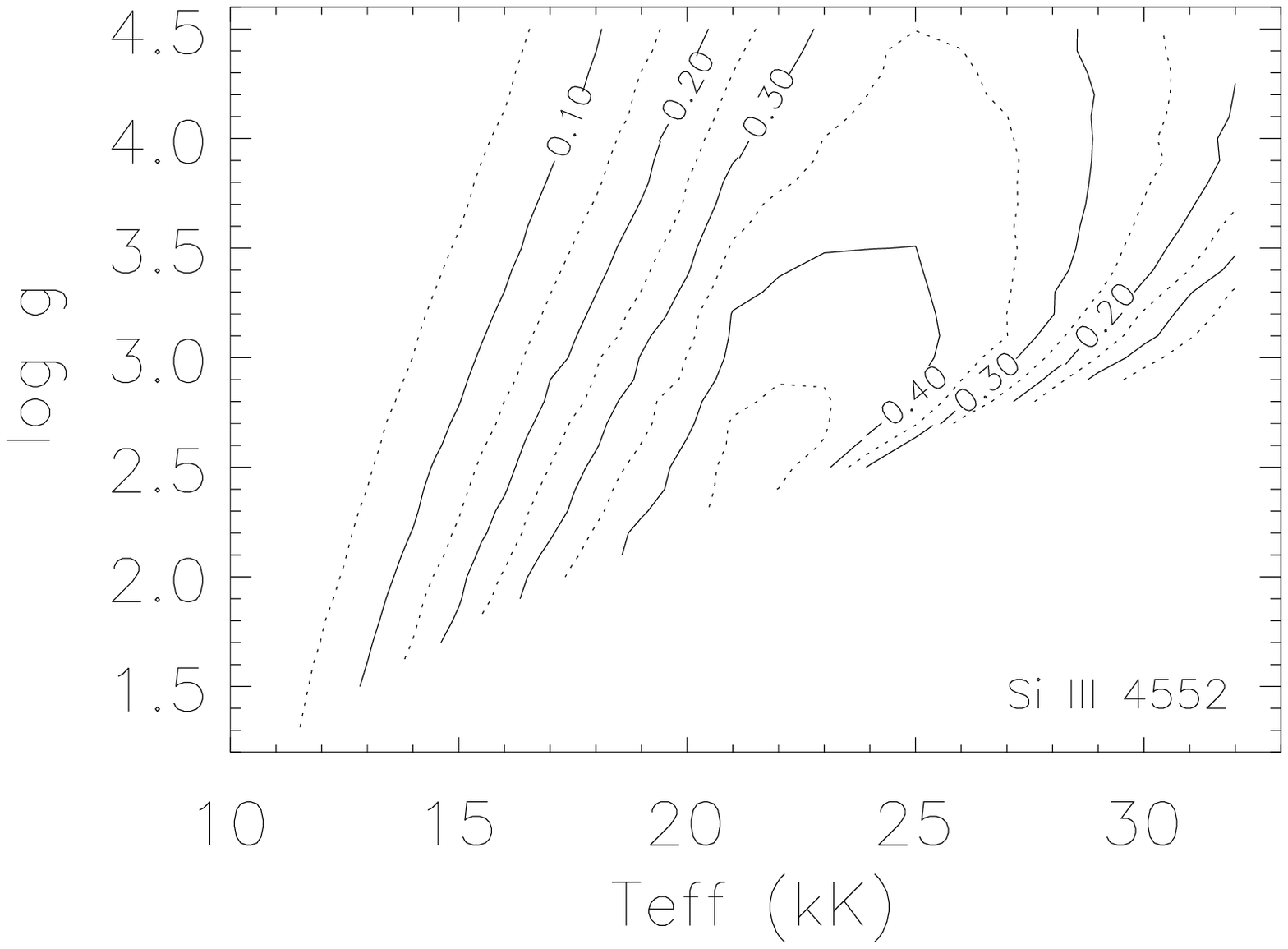}\hspace{-0.5cm}
\includegraphics[scale=0.35]{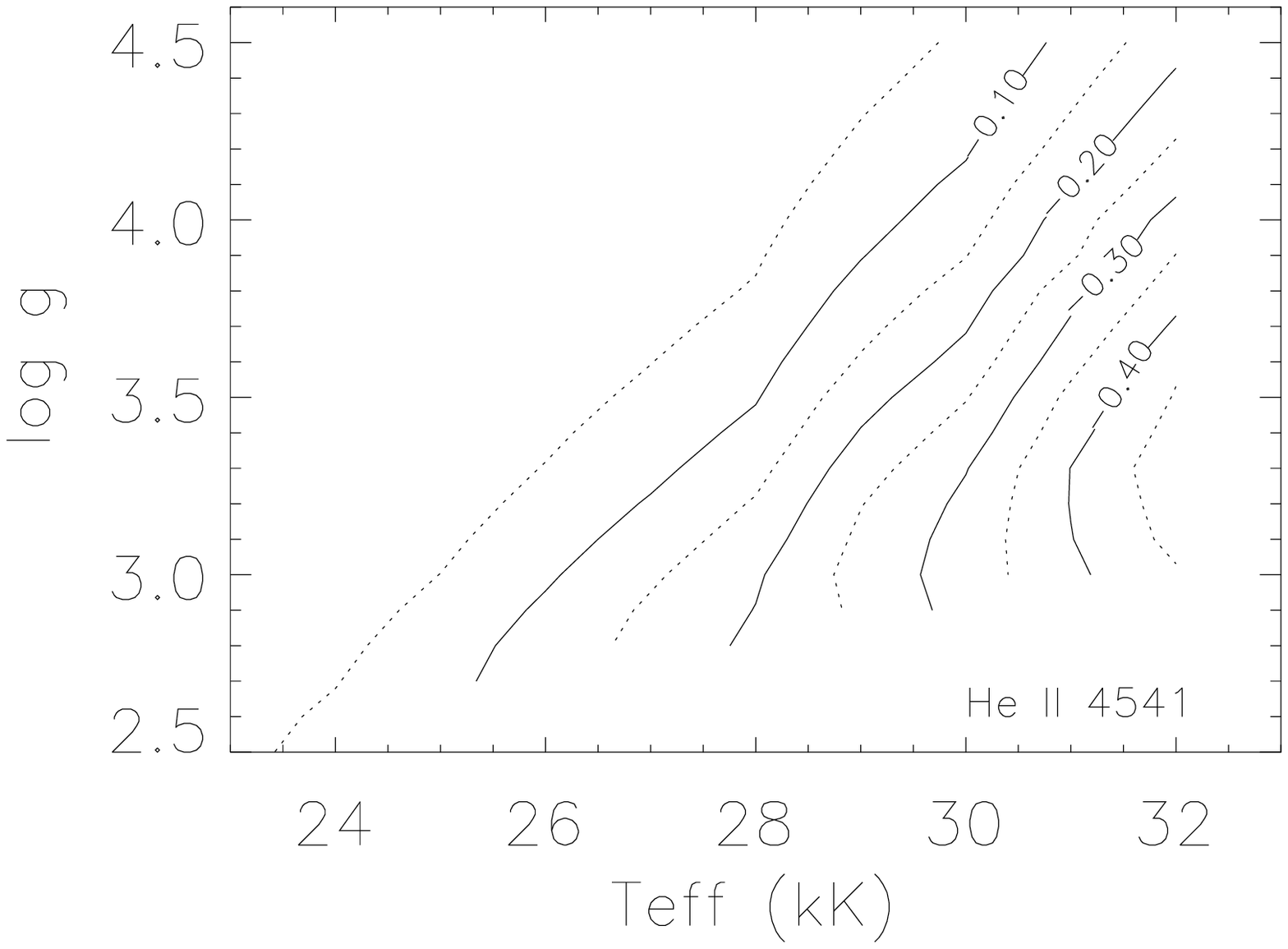}\\
\includegraphics[scale=0.35]{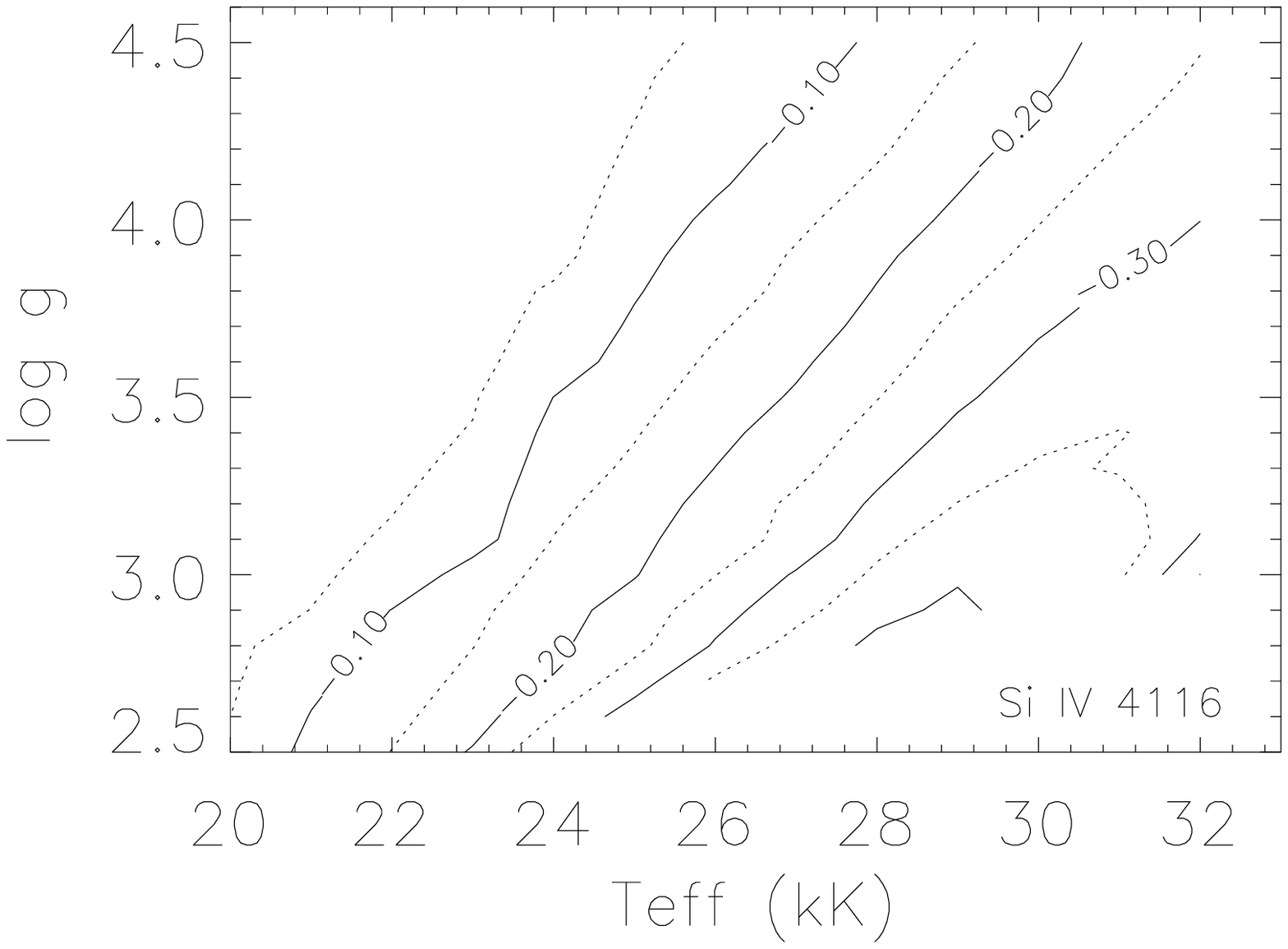}\hspace{-0.5cm}
\includegraphics[scale=0.35]{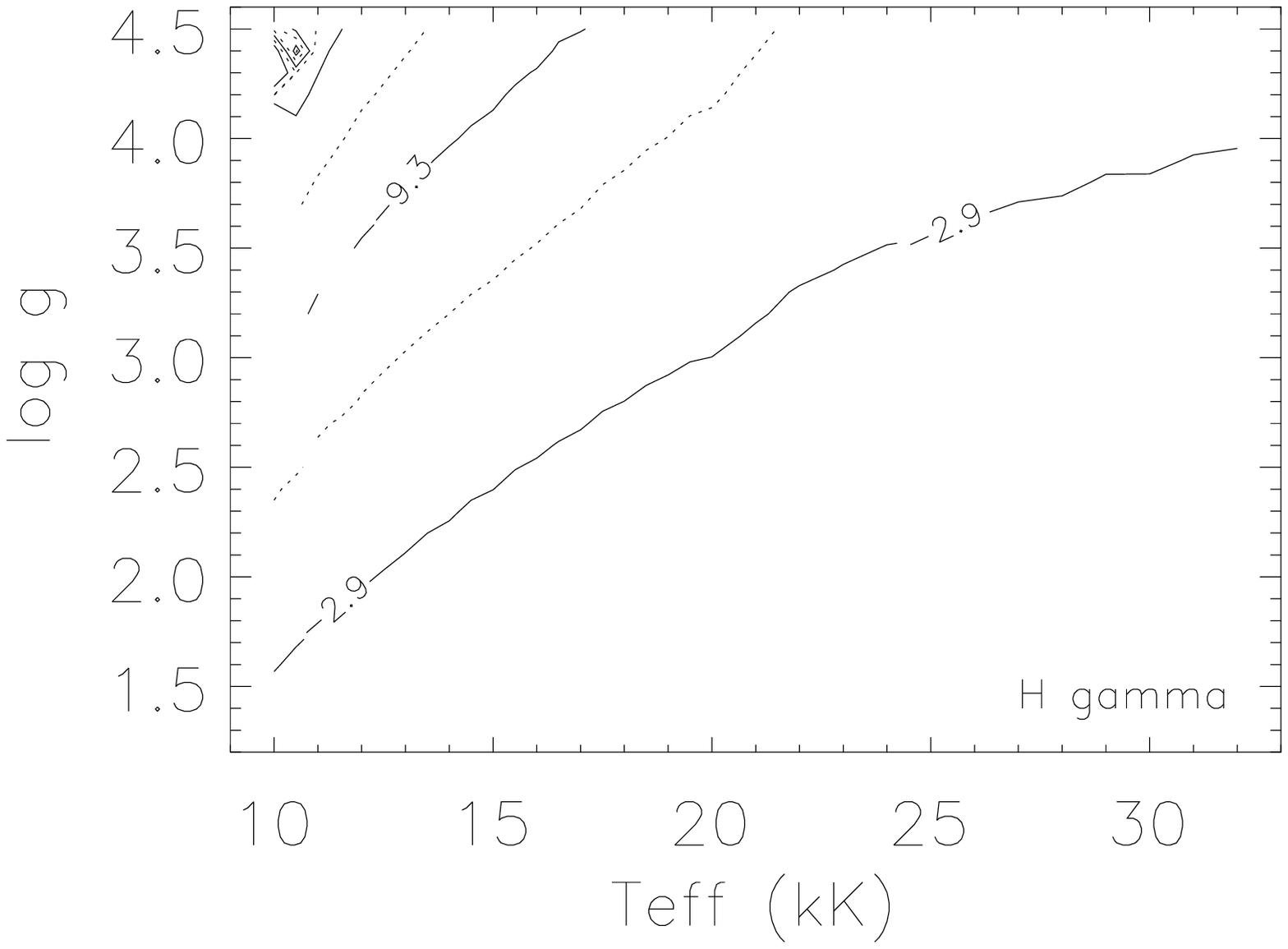}\\
\caption{Isocontours in the ($T_{\rm eff}$, log $g$) plane for several
diagnostic lines. Isocontours shown are for weak winds only.
Left panel: Si~II~4128 (top), Si~III~4552 (middle) and Si~IV~4116 (bottom).
Right panel: He~I~4471 (top), He~II~4541 (middle) and H$\gamma$ (bottom).}
\label{iso}
\end{figure}

\begin{figure}[!ht]
\includegraphics[scale=0.35]{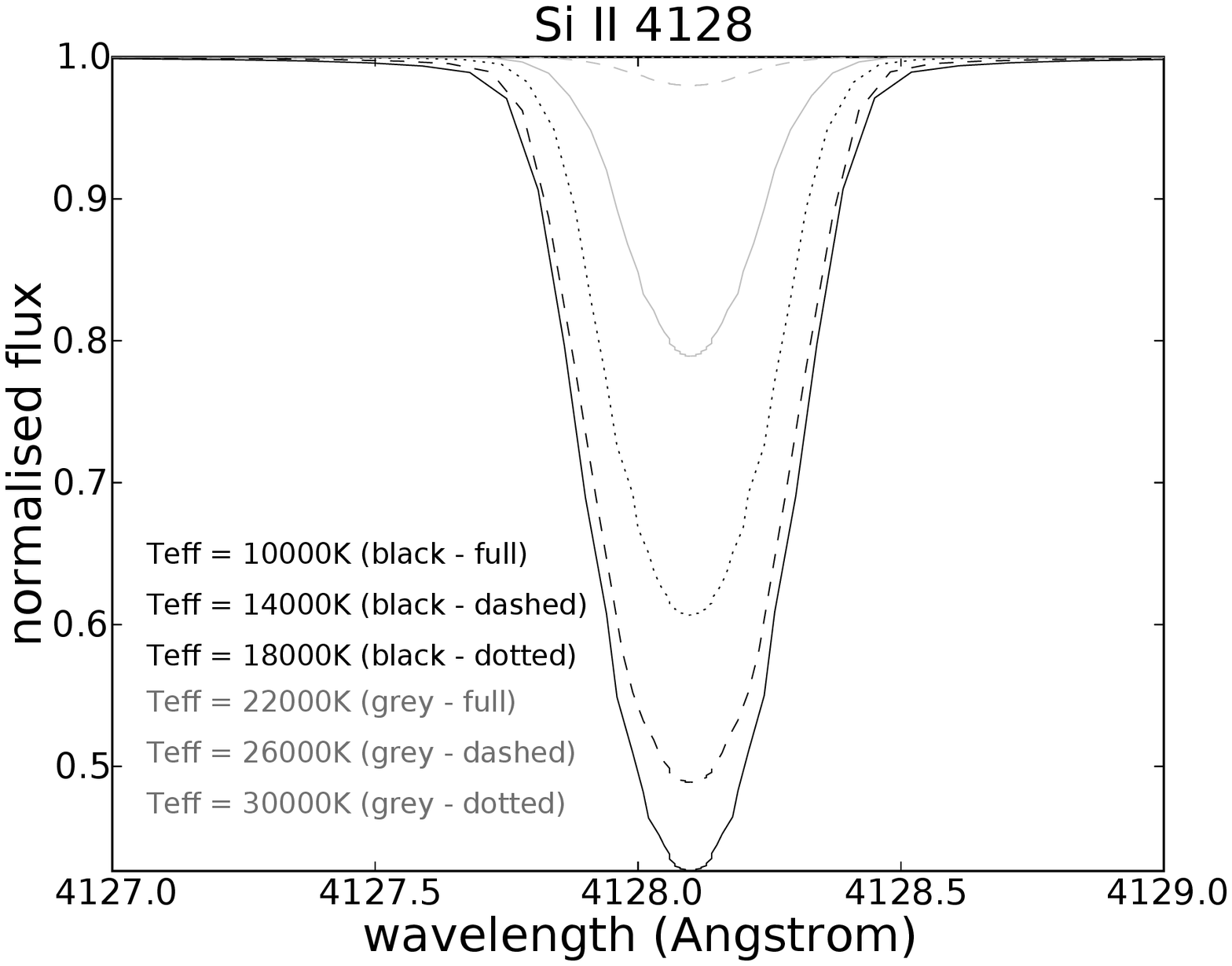}
\hspace{-0.8cm}
\includegraphics[scale=0.35]{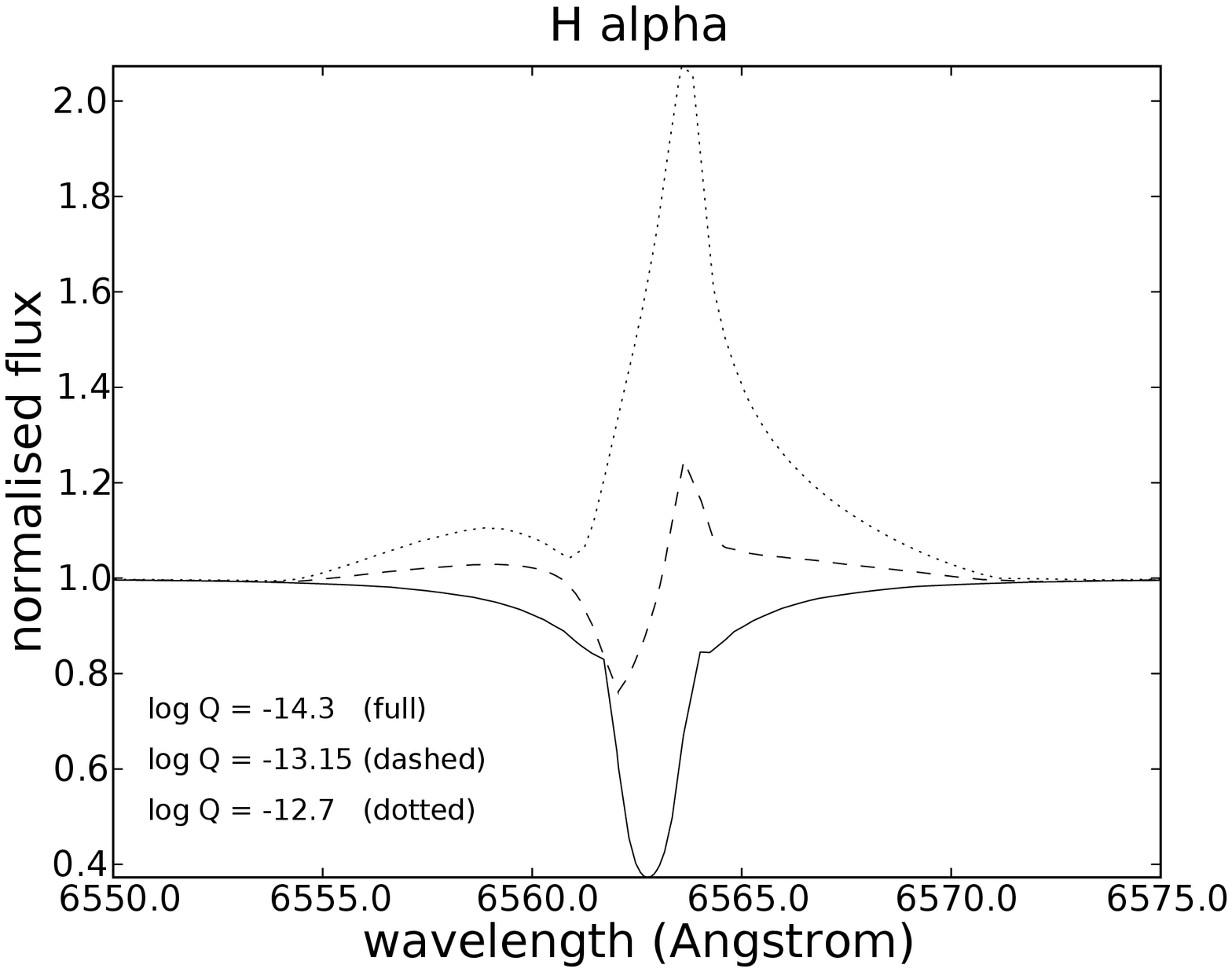}\\
\includegraphics[scale=0.35]{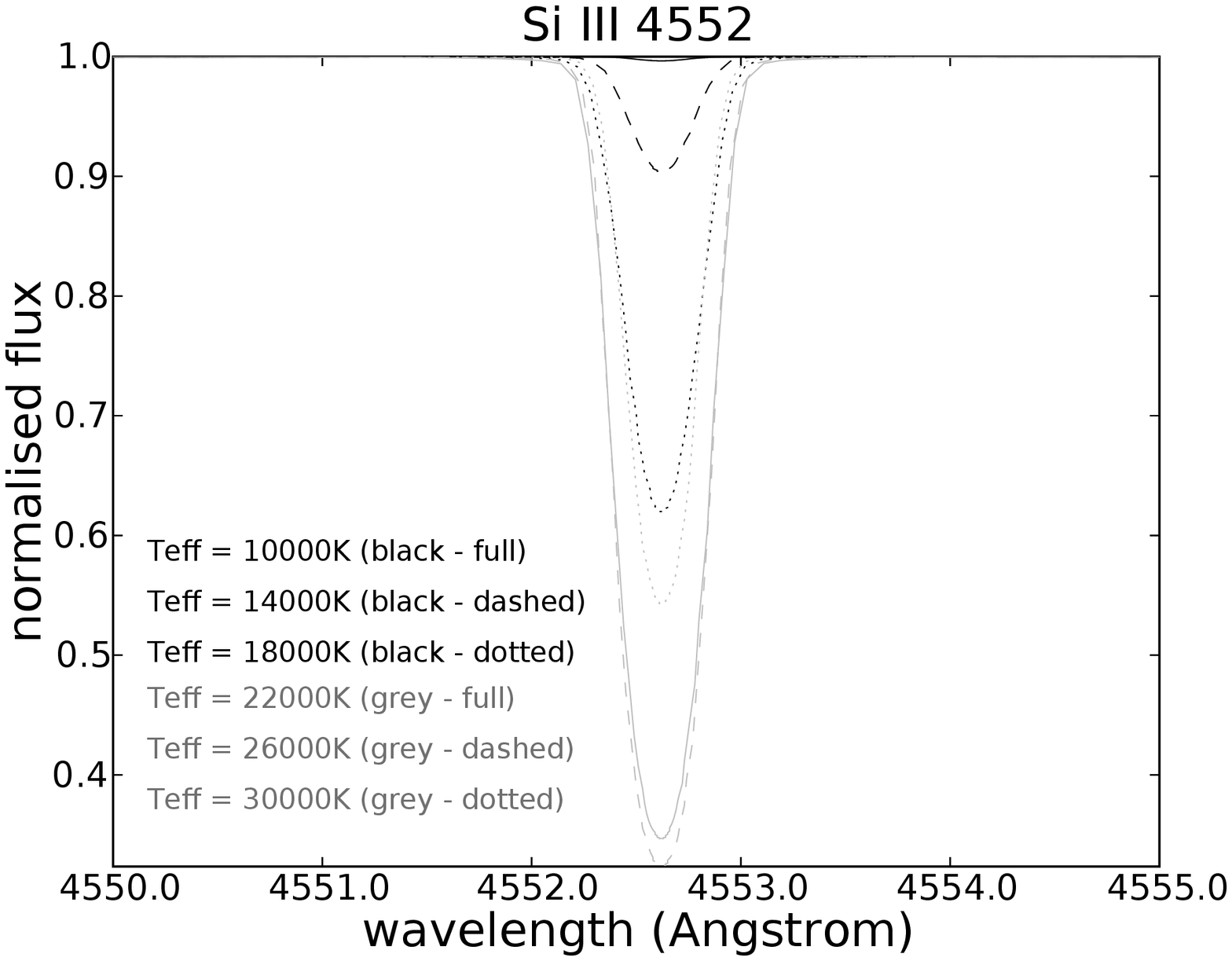}
\hspace{-0.8cm}
\includegraphics[scale=0.35]{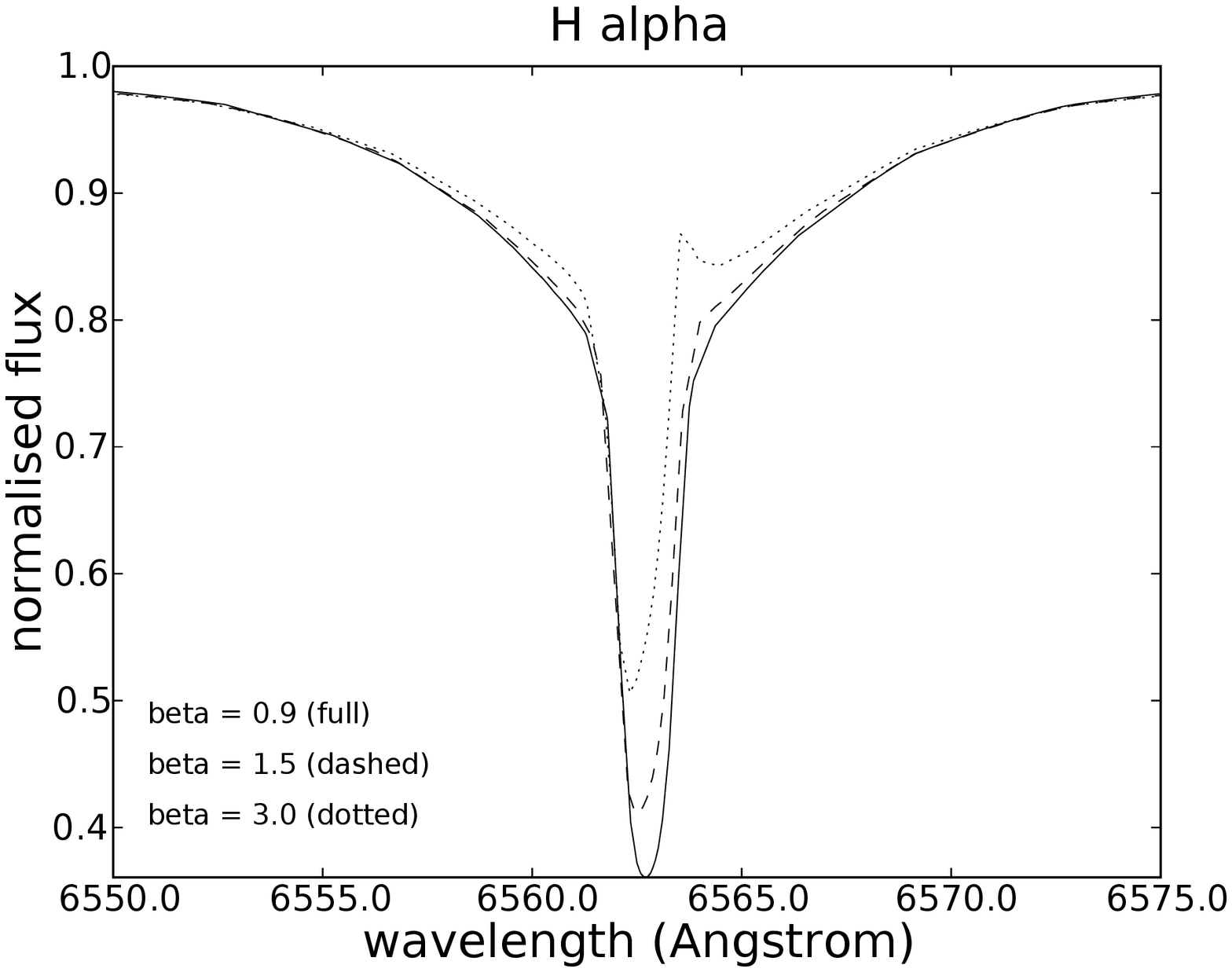}\\
\includegraphics[scale=0.35]{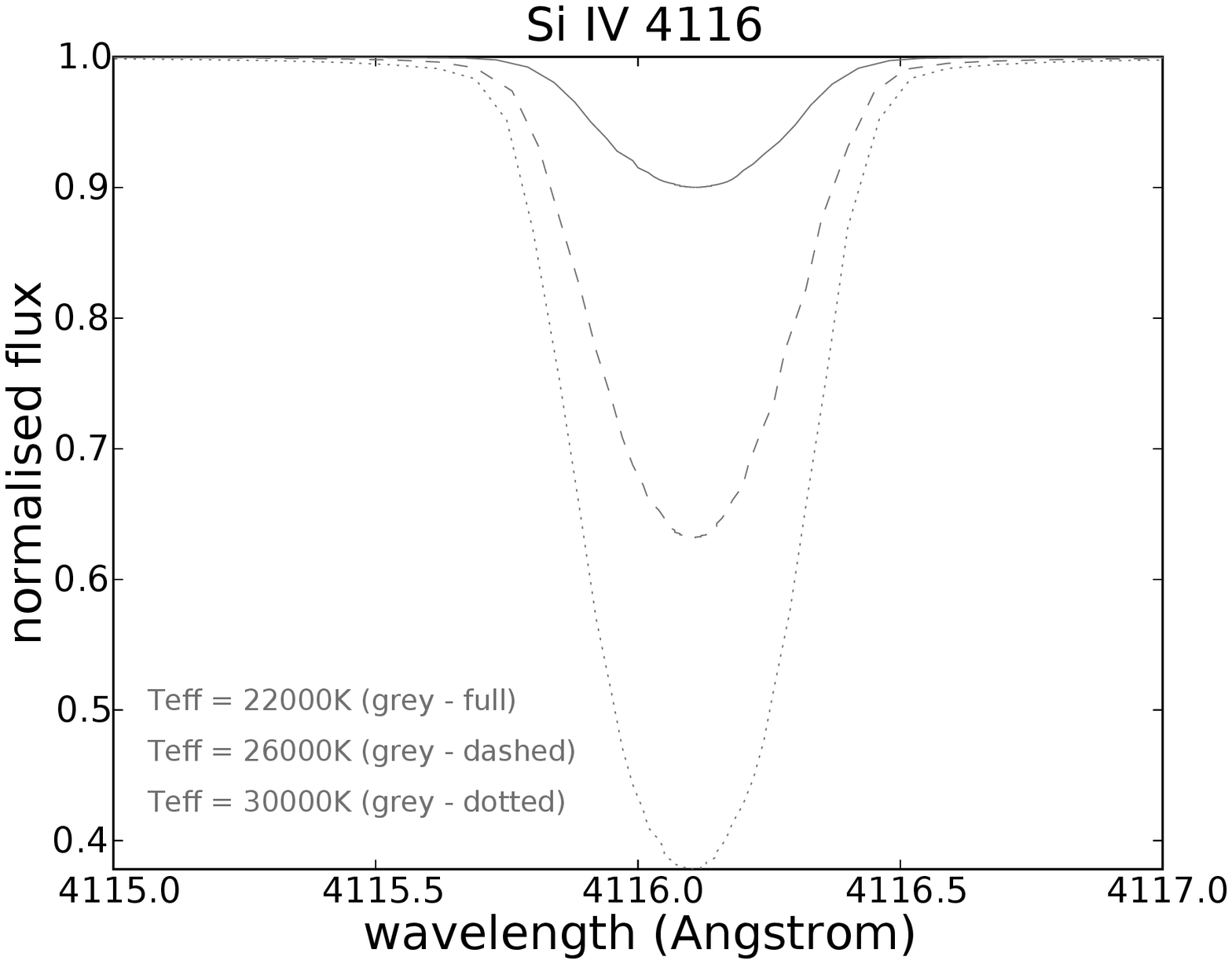}
\hspace{-0.8cm}
\includegraphics[scale=0.35]{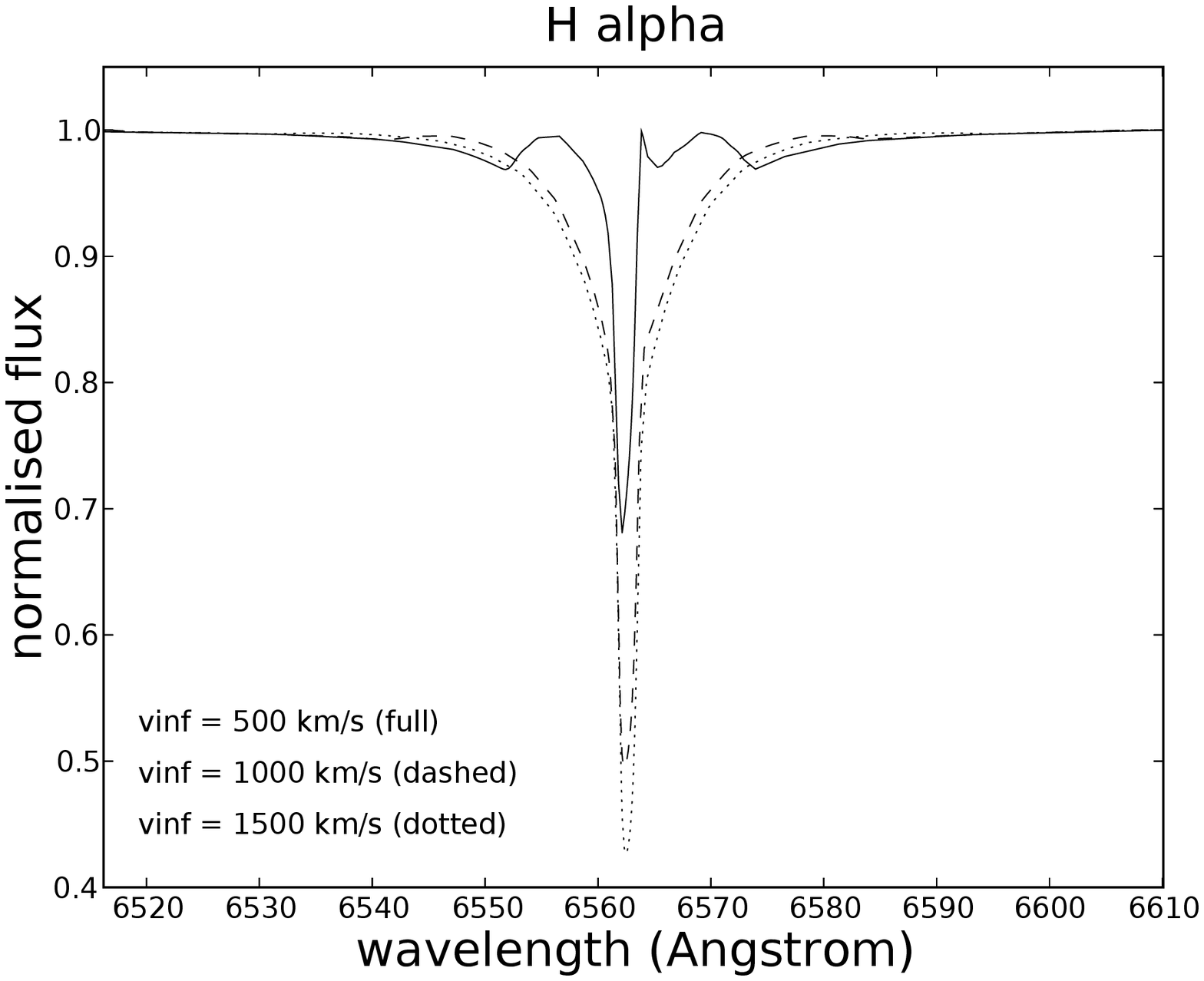}\\
\caption{A simplified, but representative, picture of the effects of $T_{\rm
eff}$ on the Si~II, III and IV lines (left panels) and the wind properties log
$Q$, $\beta$ and $v_{\infty}$ on the H$\alpha$ profile (right panels).}
\label{features1}
\end{figure}

\begin{figure}[!ht]
\includegraphics[scale=0.35]{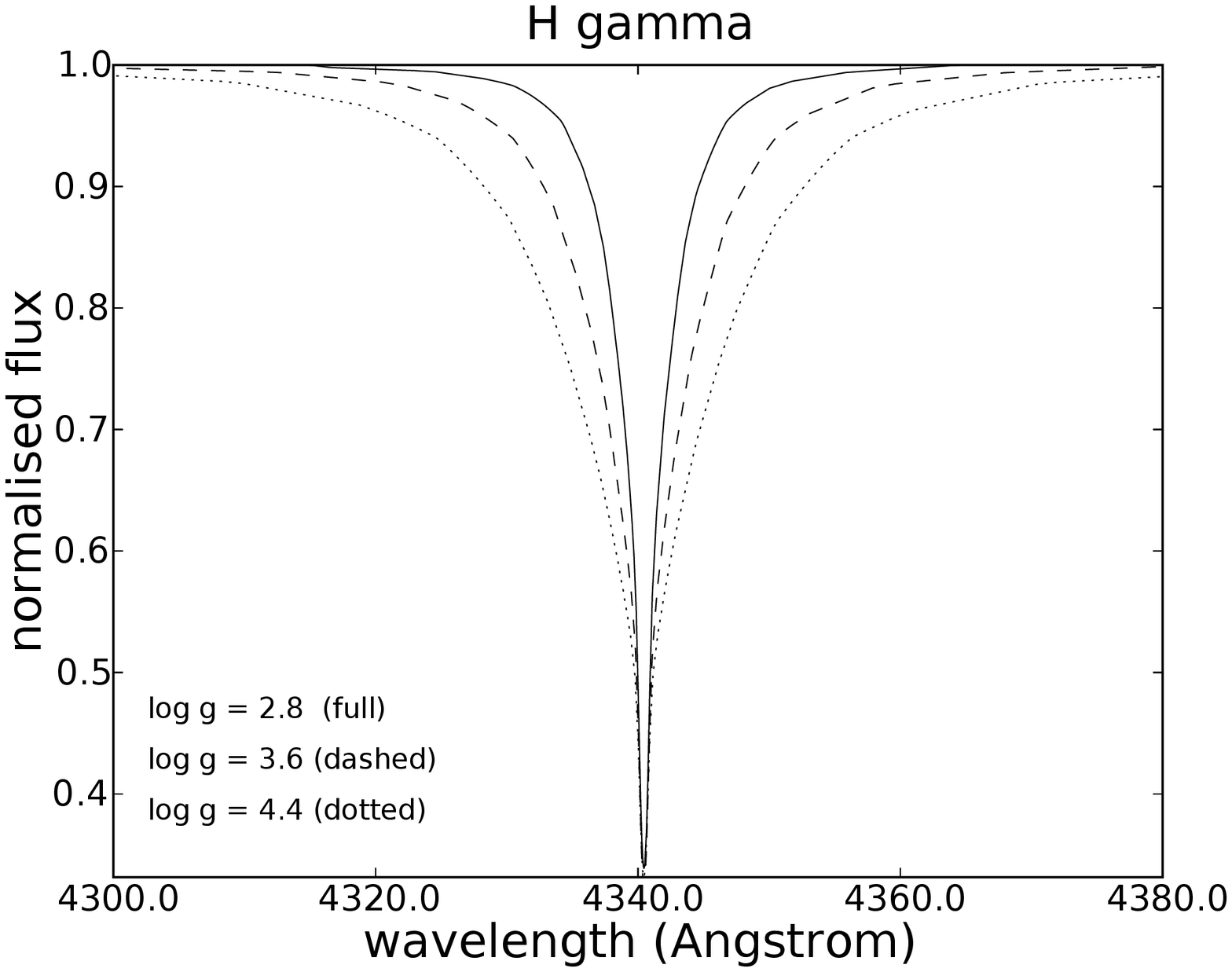}
\hspace{-0.8cm}
\includegraphics[scale=0.35]{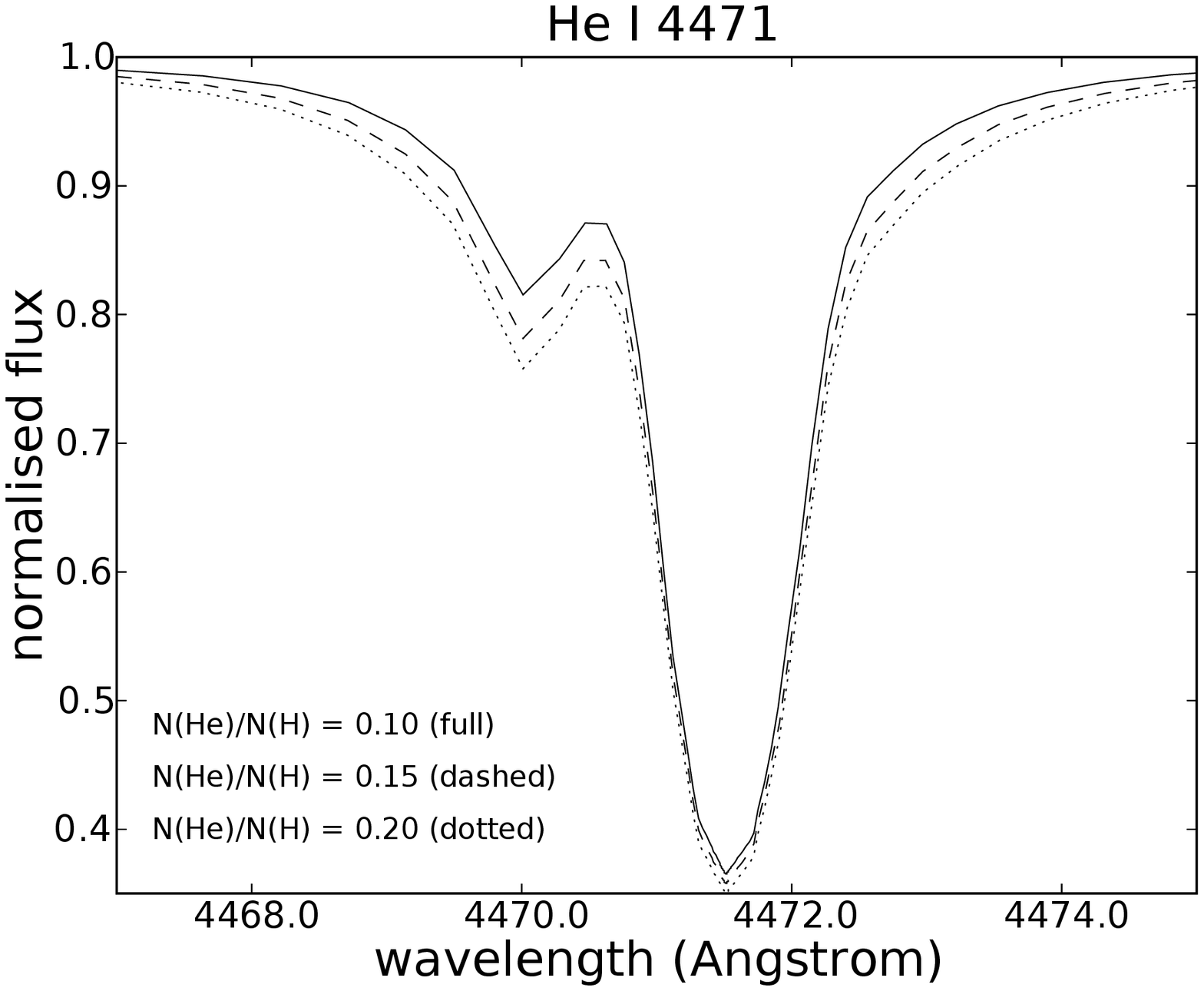}\\
\includegraphics[scale=0.35]{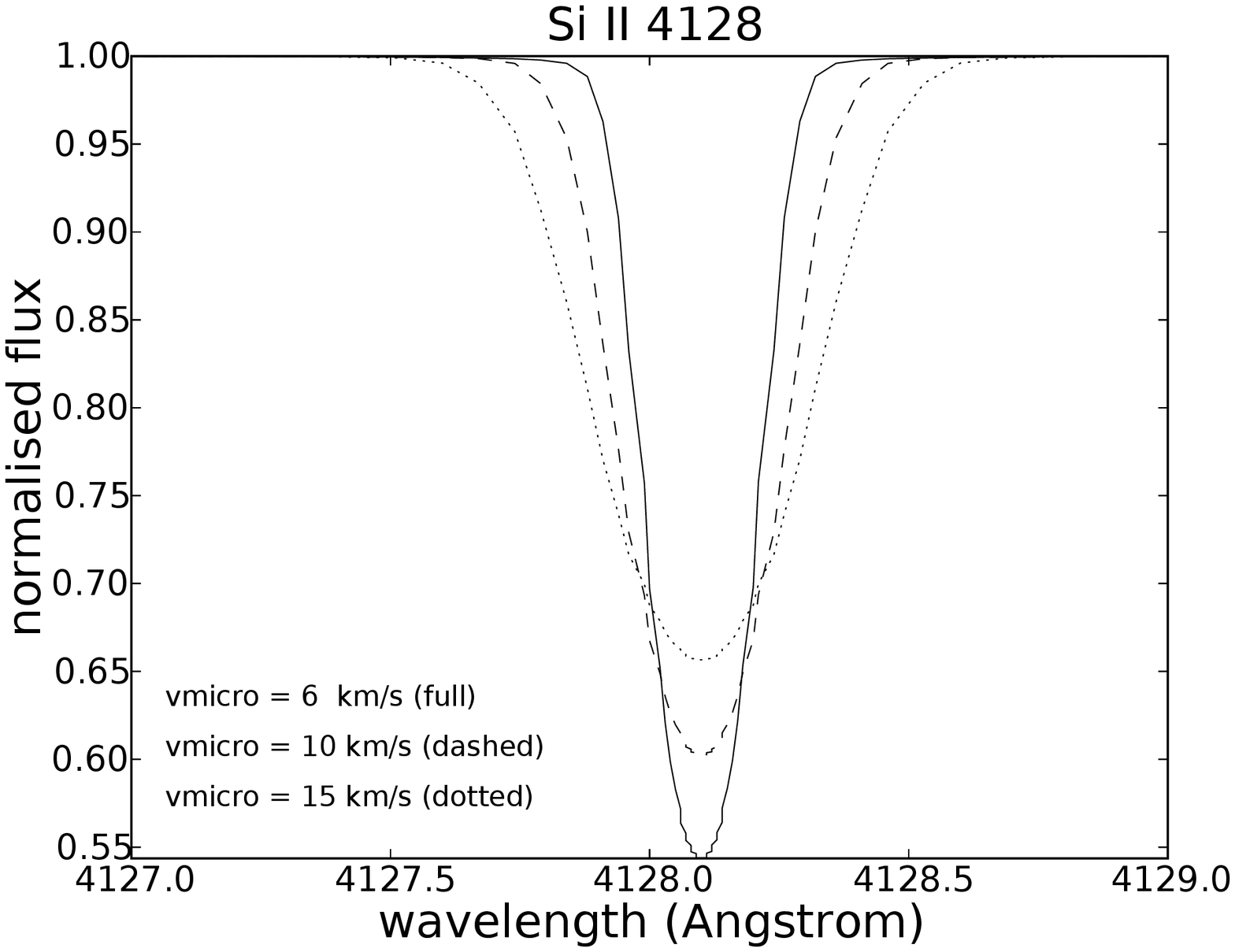}
\hspace{-0.8cm}
\includegraphics[scale=0.35]{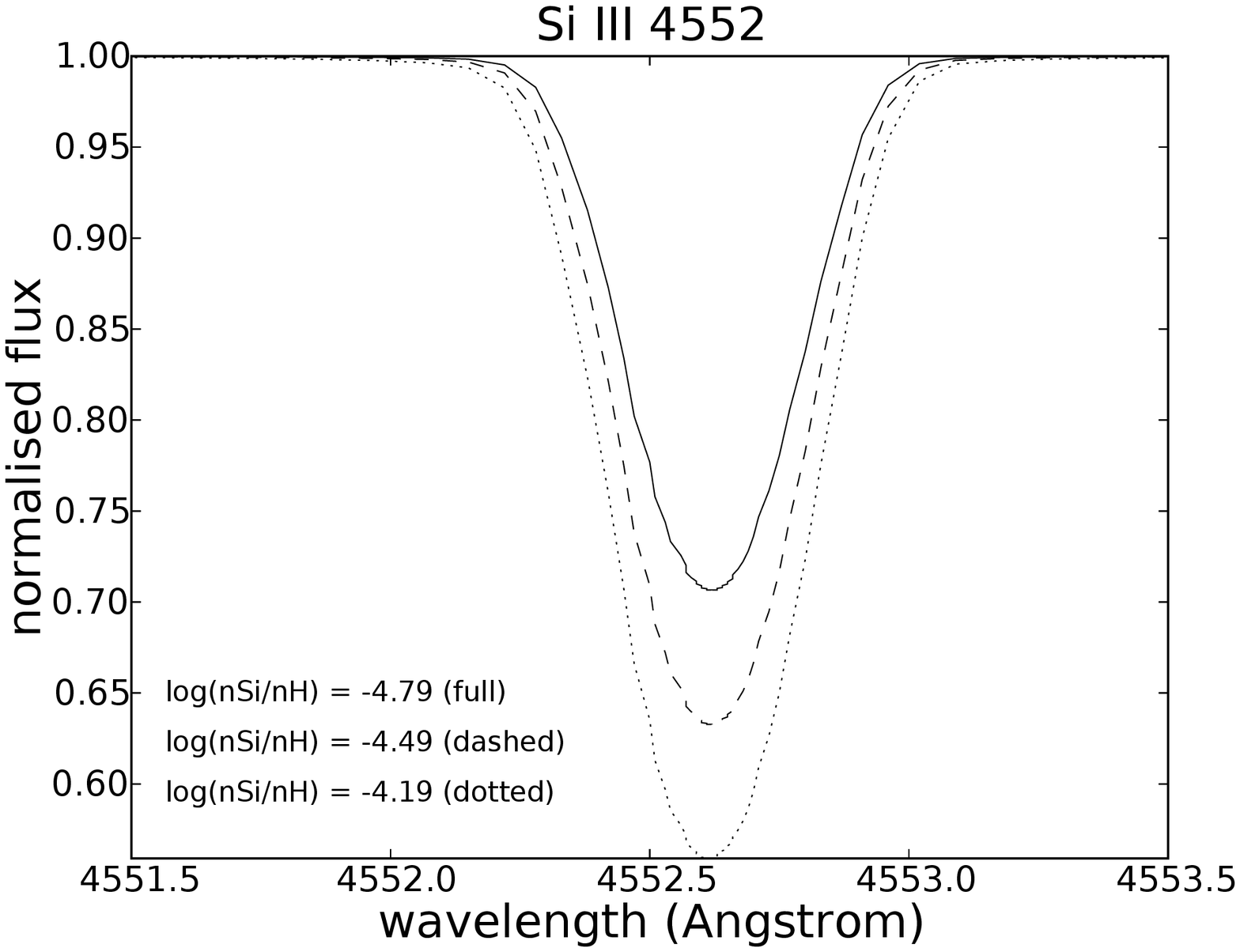}\\
\caption{A simplified, but representative, picture of the influence of the
gravity on the wings of H$\gamma$ (left panel: top), of the microturbulent
velocity on Si~II~4128 (left panel: bottom) and the effect of changing the
He and Si abundance (right panels).}
\label{features2}
\end{figure}

Fig.\ref{iso} shows the isocontours for the equivalent width (EW) of
some selected lines. They are based on the complete model grid and
represent the dependence of each line on both the effective temperature and
the gravity. 

Si lines have been chosen because of their strong dependence on temperature.
When Si~II, Si~III and Si~IV are considered in parallel, they unambigously
define the effective temperature. This is demonstrated in the left panels of
Fig.\ref{iso}, by the isocontours for a representative Si~line for each of the
three ionisation stages: Si~II~4128, Si~III~4552 and Si~IV~4116. Increasing the 
temperature in the lower temperature region results in an opposite effect
regarding the EW of Si~II and Si~III. Whereas, in the higher temperature
region, a similar effect can be observed between Si~III and Si~IV. This
behaviour is demonstrated directly in the line profiles shown
in the left panels of Fig.\ref{features1}.

The same is true for He. In the low temperature region, where He~I becomes
insensitive to gravity, it is the most sensitive diagnostics for changes in
the effective temperature (note the almost vertical isocontour). At higher
temperatures, where He~I looses its diagnostic sensitivity with respect to
$T_{\rm eff}$, He~II takes over (see Fig.\ref{iso} - right panel - top and
middle). In these regions He~I remains useful, but now as a good gravity
indicator (note the almost horizontal isocontours). In this way, He~I will
provide us with a {\it second} check for the derived $T_{\rm eff}$ and log $g$
values, but, most important, will allow to derive the He-abundance (from the
absolute strength of the lines, see below).

Isocontours of H$\gamma$ (Fig.\ref{iso} - lower right panel) show that it is
equally dependent on $T_{\rm eff}$ and log $g$. 
Since we can fix $T_{\rm eff}$ already from the Si~lines, we can
easily determine the gravity by fitting its wings, since these are most
affected by even small changes in surface gravity (see Fig.\ref{features2}).
Thus, H$\gamma$ is our main gravity indicator.

To obtain the best possible information about the wind properties, we need a
line which is sufficiently affected by changes in the wind parameters. The best
suited line is H$\alpha$, which is formed in the outer photosphere and
lower/intermediate wind. In the right panels of Fig.\ref{features1}, we show
the influence of the wind strength parameter log $Q$, the velocity-field
exponent $\beta$ and the terminal wind velocity v$_{\infty}$ on the shape of the
resulting (P Cygni) profile.

Of course, this is just the easy picture. There are more things which
influence the line profiles, in particular microturbulent velocity v$_{\rm
micro}$ and abundance (He and Si). Their effects are shown in
Fig.\ref{features2}, and we will note here only that both quantities have to
be derived in parallel, by a rather complex procedure, which nevertheless can
be automatized as well.

For reason of brevity, we have restricted ourselves to these figures,
because they show the sensitivity to the most fundamental parameters of the
star. More could have been drawn, such as the influence of the
rotational velocity and the macroturbulent velocity v$_{\rm macro}$, which
enter the (synthetic) profiles by simple convolutions.

\section{Discussion}

We have computed a dense grid of NLTE atmosphere models with and without winds
covering all spectral types B. The computations were performed continuously 
during a period of seven months on a dedicated Linux cluster with 20 CPUs (3800
MHz processors with 4 Gb RAM memory and 8 Gb Swap memory).  Whenever available
these 20 CPUs were extended with 40 extra CPUs (among which 8 more of 3800MHz
and 32 of 3400MHz).

In order to avoid unnecessary repetition of the computation of such a huge
grid, we will offer BSTAR06 to the community for further research in the near
future.

By translating the different effects of the individual parameters on the
selected line profiles into code, one can finally perform an automated line
profile fitting procedure. This process is currently in its phase of
development. Once the automated tool does exist, we will be able to analyse B
type stars in a fast and effective way, which hopefully leads to a better
understanding of this hot star regime. 
In particular, the interpretation of the COROT grid should lead to a much better
calibration of hot stars across the whole B type spectral range.

\acknowledgements{We thank Alex de Koter and Rohied Mokiem very heartily for 
many fruitful discussions and their very useful suggestions for the
construction of the model grid. We are also very grateful to Erik Broeders who
helped a lot with the technical aspects of the grid calculations.}

=======================================
 \question{Ma\'{\i}z Apell\'aniz}  For this grid, will you release the full 
 SEDs or just the line profiles?
 
 \answer{Lefever}  Only the H, He and Si profiles will be released. 
 More line profiles can easily be added using the calculated model structure,
 if individual users would require this.  
 
 \question{Ma\'{\i}z Apell\'aniz}  Once you remove systematic 
 uncertainties, is it possible to measure B-star temperatures with 
 optical-NIR photometry?
 
 \answer{Lefever}  As far as I know, this is only true for main-sequence 
 stars (not for giants or supergiants). However, this is still based on LTE 
 models, whereas we want to include also the NLTE effects occurring in these 
 massive stars.
 
 \question{Cohen}  How do you account for the broadening of the hydrogen 
 lines?
 
 \answer{Lefever} Stark broadening is included to correctly treat the 
 broadening of the hydrogen lines.
 
 \question{Adelman}
 How do these models compare with the recent TLUSTY  O and B model grid?
 
 \answer{Lefever}
 TLUSTY is a plane-parallel code, whereas the FASTWIND models assume
 spherical symmetry. TLUSTY does not include any wind, whereas this is a
 prerequisite to deal with supergiants.
 
 \question{Adelman}  I agree with Dr. Ma\'{\i}z Apell\'aniz that one can 
 get good effective temperatures and surface gravity results for 
 main-sequence band mid- to late-B stars. The use of NLTE model atmospheres 
 is necessary when NLTE effects influence the continuum.
 
 \answer{Sterken}  Sure, but this work deals with massive stars way above 
 the main sequence.

 =======================================

\end{document}